\documentstyle[12pt,amsfonts,amssymb,epsf]{article}

\begin{document}

\def\be{\begin{equation}}
\def\ee{\end{equation}}
\def\ba{\begin{eqnarray}}
\def\ea{\end{eqnarray}}
\def\P{\Bbb{P}}
\def\R{\Bbb{R}}
\begin{center}
{\bf Does the ``Soft Pomeron'' Cope with the  HERA Data?}

\vspace*{0.2cm}
V.A.Petrov\footnote{E-mail: petrov@mx.ihep.su},
A.V.Prokudin\footnote{E-mail: prokudin@th.ihep.su}\\

\vspace*{0.2cm}
{\it Division of Theoretical Physics, Institute For High Energy Physics, 142284 Protvino, Russia.}
\end{center}

\vspace*{0.4cm}
%\centerline{Abstract}

\vspace*{0.2cm}
{\small 
It is demonstrated that the "soft pomeron" (together with a secondary reggeon)
describing cross-sections of hadron-hadron collisions is fairly able to
describe the energy dependence of cross-sections of exclusive (virtual)
photoproduction of vector mesons $(\rho^0, \varphi, J/\psi)$ obtained at
HERA. However these poles are insufficient for description of the total DIS
cross-section. 
}

\vspace*{1cm}
The effect of acceleration of the cross sections growth with c.m.s. energy
$\sqrt{s}$ in presence of
a hard scale ( a current virtuality, a high mass in the final state etc) 
revealed at HERA~[1] raises  a question on validity ( or sufficiency) of
the Regge-pole model that has a good reputation in the description of cross sections of
hadron-hadron scattering~[2]. This phenomenon can be described by a simple power-like 
phenomenological parametrization:    

\be
\sigma_{\mbox\small{{tot}}}^{\gamma^*p}\sim s^\lambda,\;\;\;
\lambda \geq 0.2
\ee
where the exponent $\lambda$ not only exceeds its ``soft'' counterpart $\Delta=\alpha_{\Bbb
{P}}(0)-1\simeq 0.08$ in the Regge-pole method but also depends on
the current virtuality  $Q^2$. In the framework of the complex J-plane
(in the $t$-channel) this means that there are some singularities located to the right
of $Re J=\alpha_{\P}$ and which depend on outer kinematic parameters beyond $t$ so that
these poles are not of Regge type.    

If such poles exist then one should explore their nature and physical
meaning. Besides one should find out and verify restrictions 
due to general principles (unitarity in the first rate).  

 The effect discovered at HERA caused a vivid response of theoreticians.
Among numerous phenomenological theories~[3] there exists a particular method based 
on well known school~[4] of finding solutions of some specific approximations to 
Bethe-Salpeter equation for gluonic amplitudes. In fact one has to know how 
QCD works at large distances (``confinement region'') in order to work out a full
analysis, but as is well known there is still no theory answering this question.
If one adds some assumptions about behaviour in this region to perturbative
analysis then one of the practical results is prediction of Regge poles with
high intercepts $(\alpha(0)-1>0.4)$. An evident inconsistency of these poles with
observed behaviour of hadron-hadron cross sections was explained by smallness
of corresponding residues~[4]. However a ``pomeron'' of high intercept
(so called ``hard pomeron'') could be relevant for description of the cross sections 
 rapid growth revealed at HERA. It is sufficient to assume that in presence
of additional ``hard scale'' (photon's virtuality) corresponding residues are 
not small. 

Taking into account the fact that the "hard pomeron" status as a QCD solution is
based on additional hypotheses and its intercept can be restricted by unitarity~[5] it could be extremely worth trying to analyse the possibility
of using the ``soft pomeron''.

An alike problem was addressed in paper~[6]. However the solution
is not satisfactory because of a great number of parameters and presence
of additional factors of non-Regge type energy dependence.  

 In this paper we consider that amplitude for vector meson production
in the process

\begin{figure}[h]
\vskip  -1cm
\hskip 5cm \vbox to 6cm {\hbox to 6cm{\epsfxsize=6cm\epsffile{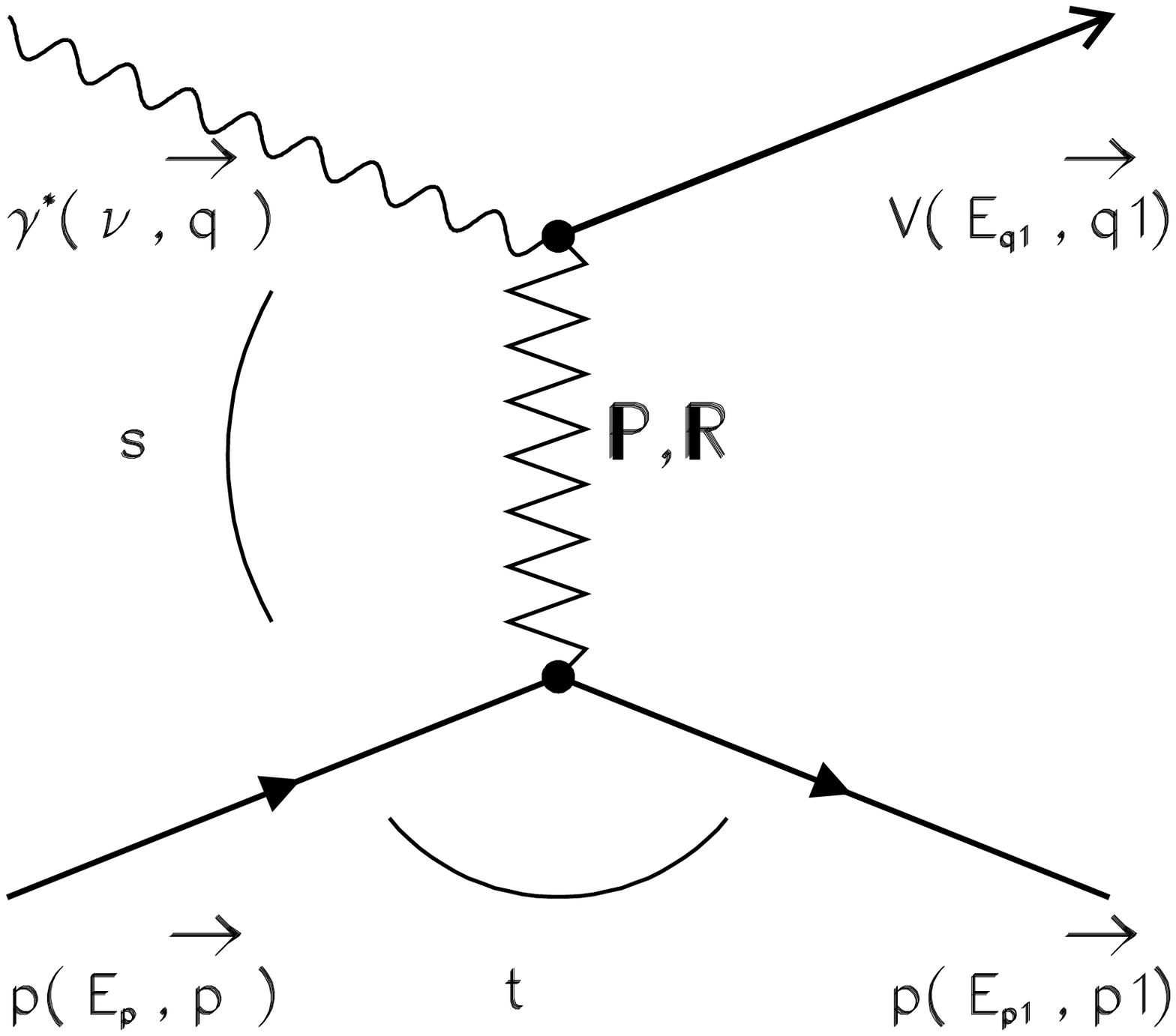}}}
\vskip  1cm
\end{figure}

$$
\gamma^*+p\to V+p\;\;\;\; 
$$
is
$$
T_{\gamma^*p\to Vp}(s,t,Q^2)=C_{\Bbb{P}} (t,Q^2)
\left (i + \cot\frac{\pi\alpha_{\Bbb{P}}(t)}{2}\right)
\left (\frac{s}{Q^2+m^2_V}\right )^{\alpha_{\Bbb{P}}(t)}+
$$
\be
C_{\Bbb{R}}(t, Q^2)
\left (i + \cot\frac{\pi\alpha_{\Bbb{R}}(t)}{2}\right)
\left (\frac{s}{Q^2+m^2_V}\right )^{\alpha_{\Bbb{R}}(t)}.
\ee
Here $Q^2$ is the photon's virtuality, $s=W^2=(p+q)^2, \;\;\Bbb{P}$ stands for the pomeron, 
$\Bbb{R}$ 
for a secondary reggeon, Regge trajectories have the following form  in a linear approximation
$$
\alpha_{\Bbb{P,R}} (t)=\alpha_{\Bbb{P,R}}(0)+\alpha'_{\Bbb{P,R}}(0)t,
$$
$$
\alpha_{\Bbb{P}}(0)-1\equiv \epsilon=0.071\pm 0.018,\;\;\;
\alpha_{\Bbb{R}}(0)-1\equiv -\mu=0.46\pm 0.25,
$$
$$
\alpha'_{\Bbb{P}}(0)=0.25,\;\;\;
\alpha'_{\Bbb{P}}(0)=1.00.
$$
The values $\alpha_{\Bbb{P,R}}(0)$ are taken from the papers~[2,7], where a
good description of hadron-hadron processes is achieved by means of 
these Regge poles. The $Q^2$ dependence of residues $C_{\Bbb{P,R}}$ is not
fixed in the Regge approximation and we assume the $t$ dependence  in
 the form $\exp \left [\frac{1}{4}R^2_{\Bbb{P,R}}(Q^2)t\right ]$ where
$R_{\Bbb{P,R}}$ are pomeron and reggeon ``radii''.

Fig.1,2,3 show the results of the description of cross sections for the processes 
\ba
\gamma^*+p &\to& \rho^0+p\nonumber \\
\gamma^*+p &\to& \varphi+p \\
\gamma^*+p &\to& J/\psi+p\nonumber 
\ea
by  means of formula (2). 

\begin{figure}

\vskip  0cm

\hskip 0.2cm \vbox to 7cm {\hbox to 7cm{\epsfxsize=7cm\epsffile{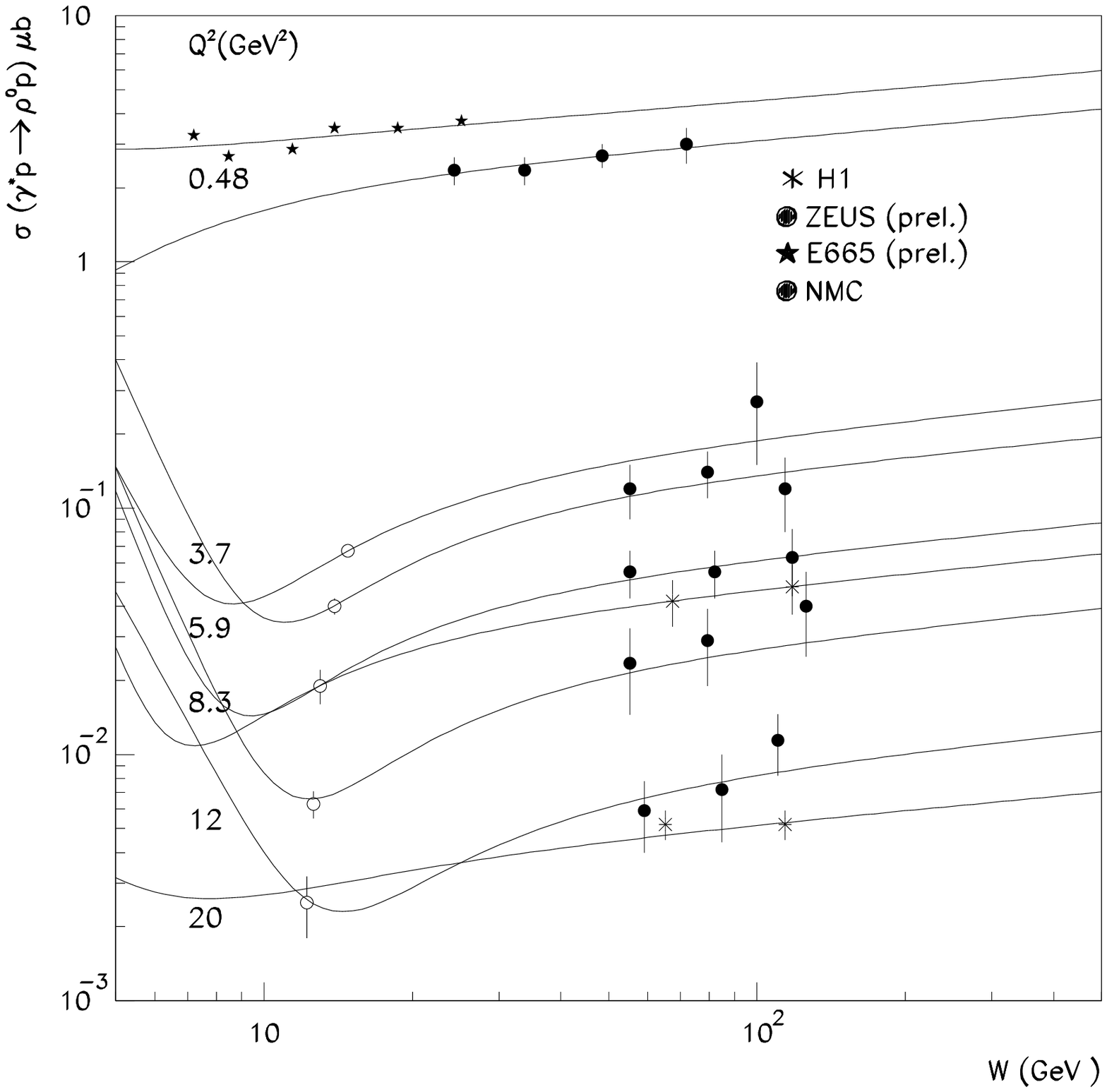}}}
\vskip  -7cm
\hskip 8.5cm \vbox to 7cm {\hbox to 7cm{\epsfxsize=7cm\epsffile{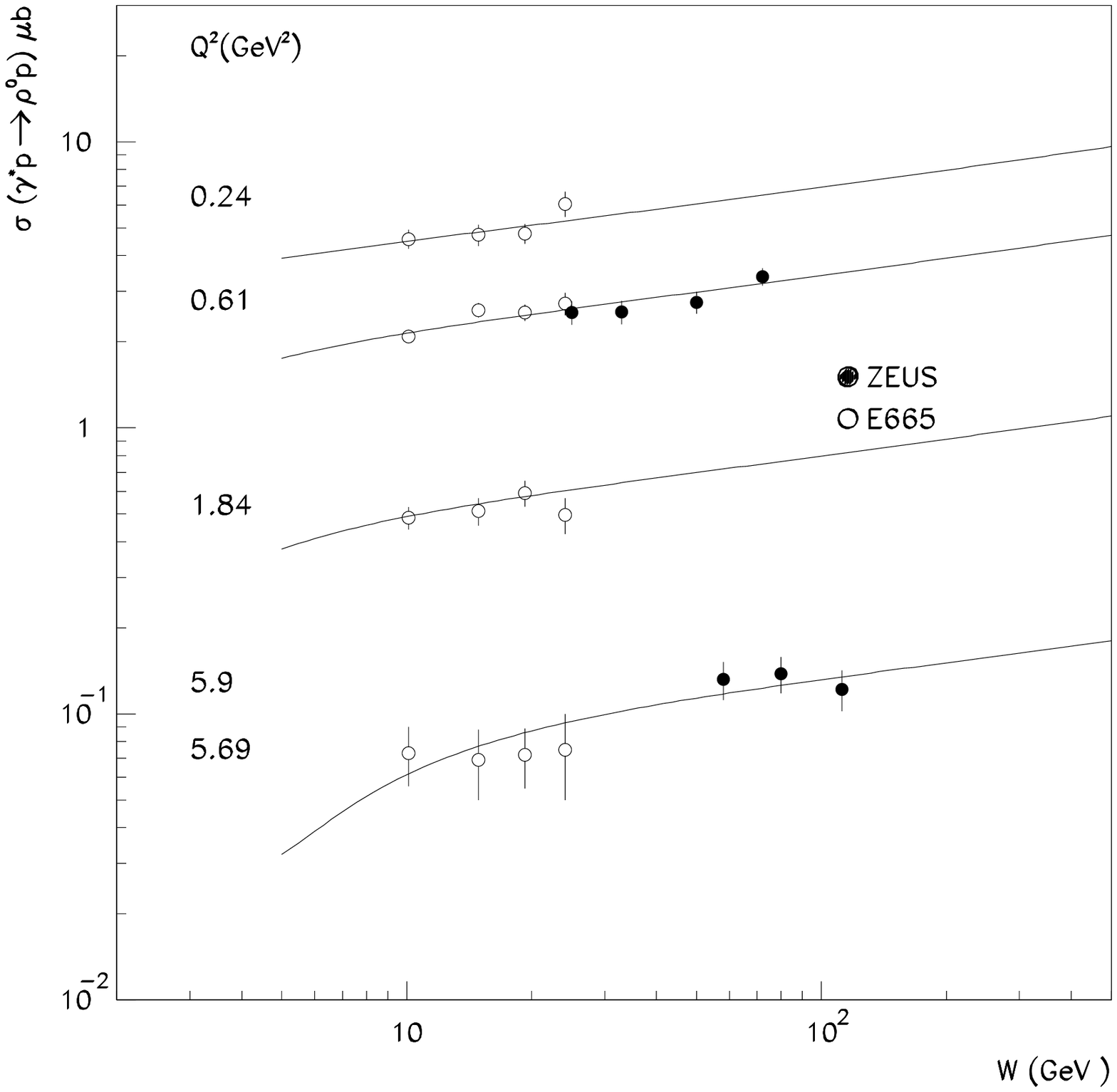}}}
\vskip 2cm
\caption{ Cross section $ \sigma _{\gamma ^* p \rightarrow
 \rho ^0 p} (W,Q^2) $  (the left picture includes experimental data from $H1$ and  $ZEUS$, the right one includes that of from $E665$ and $ZEUS$).}

\hskip 0.2cm \vbox to 7cm {\hbox to 7cm{\epsfxsize=7cm\epsffile{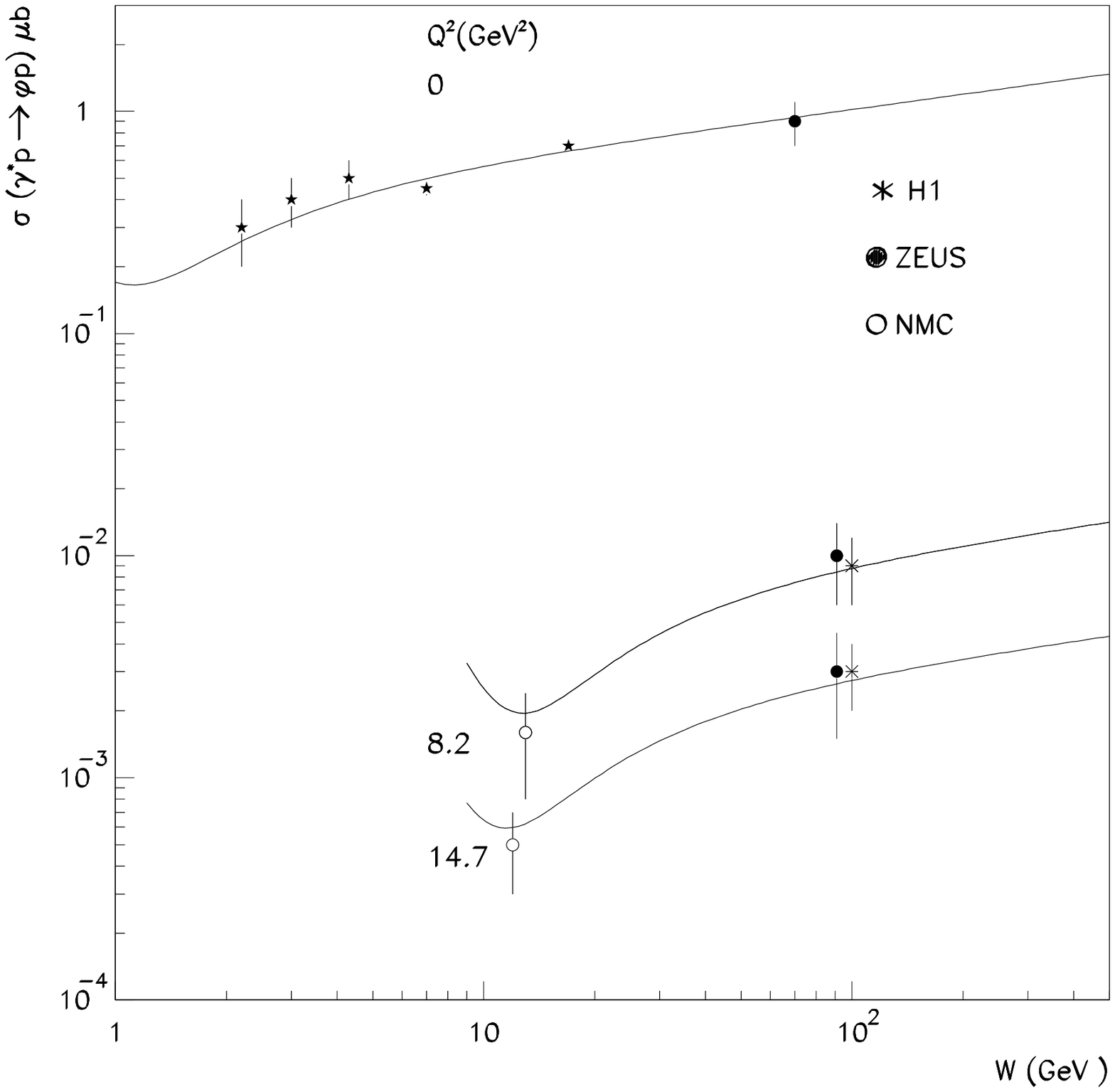}}}
\vskip  -7cm
\hskip 8.5cm \vbox to 7cm {\hbox to 7cm{\epsfxsize=7cm\epsffile{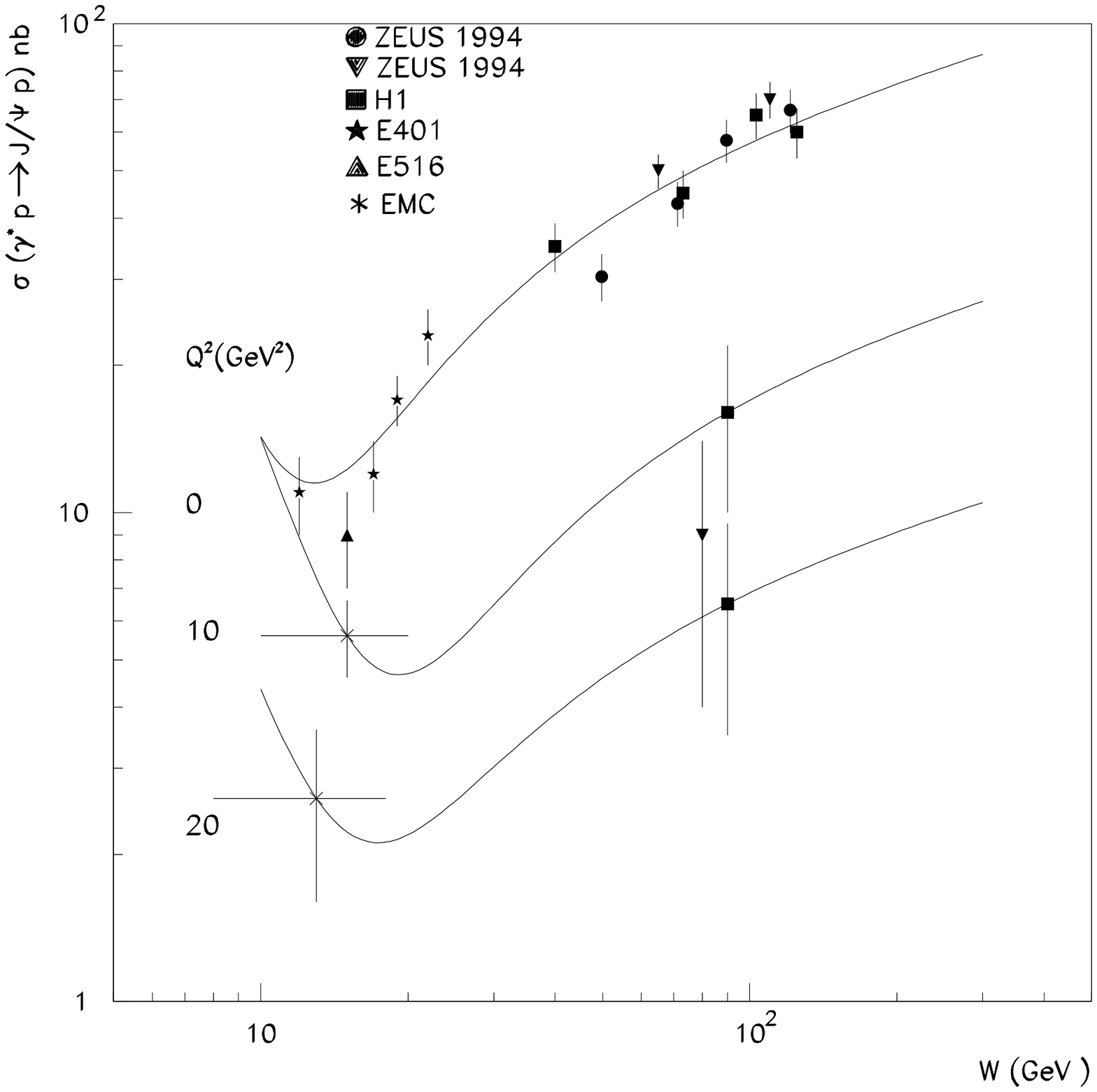}}}
\vskip 2cm
\caption{Cross section $ \sigma _{\gamma ^* p \rightarrow
 \phi p} (W,Q^2) $  (the left picture)  and  cross section $ \sigma _{\gamma ^* p \rightarrow
 J/\Psi p} (W,Q^2) $  (the right picture).}
\end{figure}

\begin{figure}
\vskip  -1cm

\hskip 0.2cm \vbox to 7cm {\hbox to 7cm{\epsfxsize=7cm\epsffile{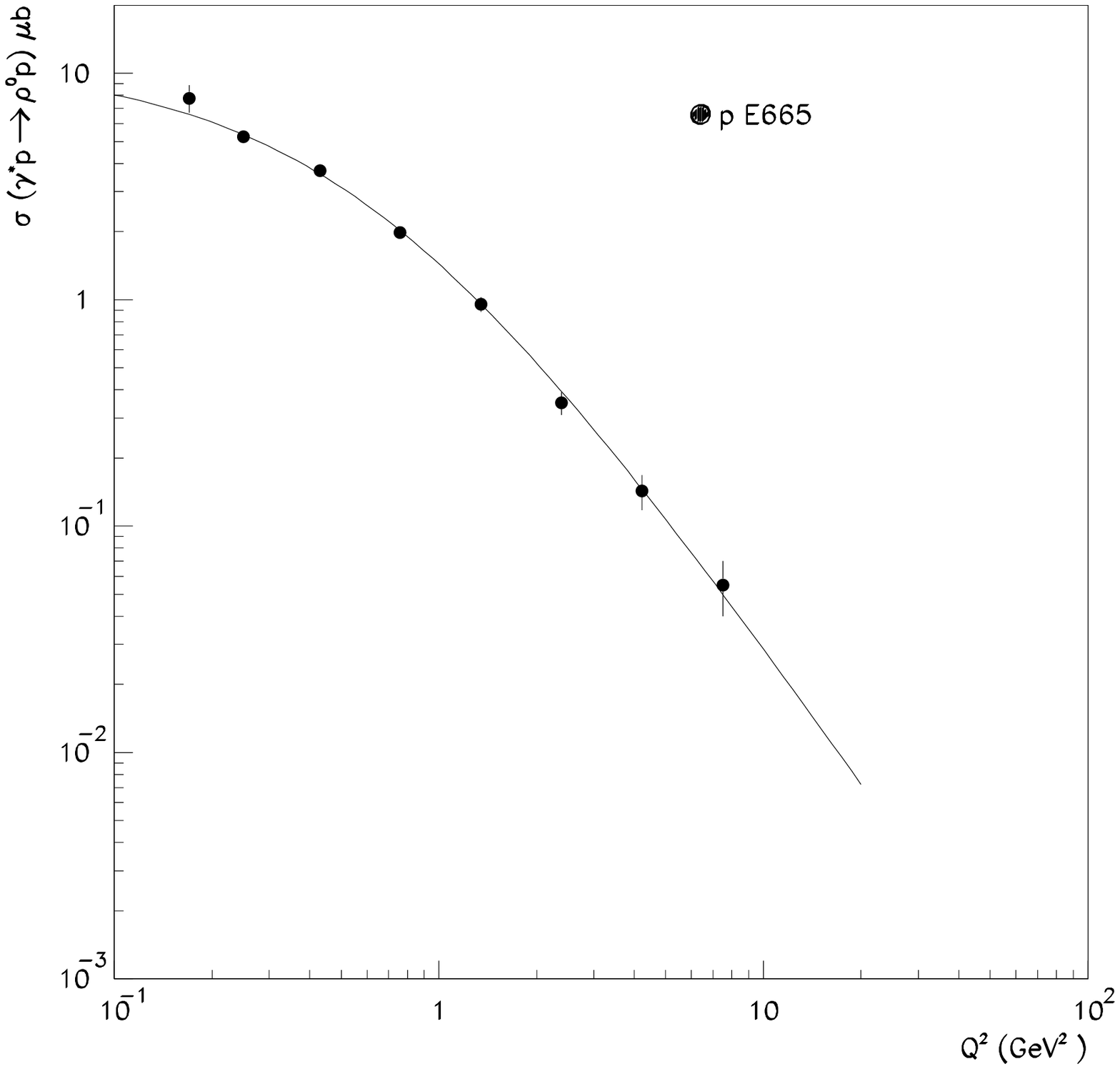}}}
\vskip  -7cm
\hskip 8.5cm \vbox to 7cm {\hbox to 7cm{\epsfxsize=7cm\epsffile{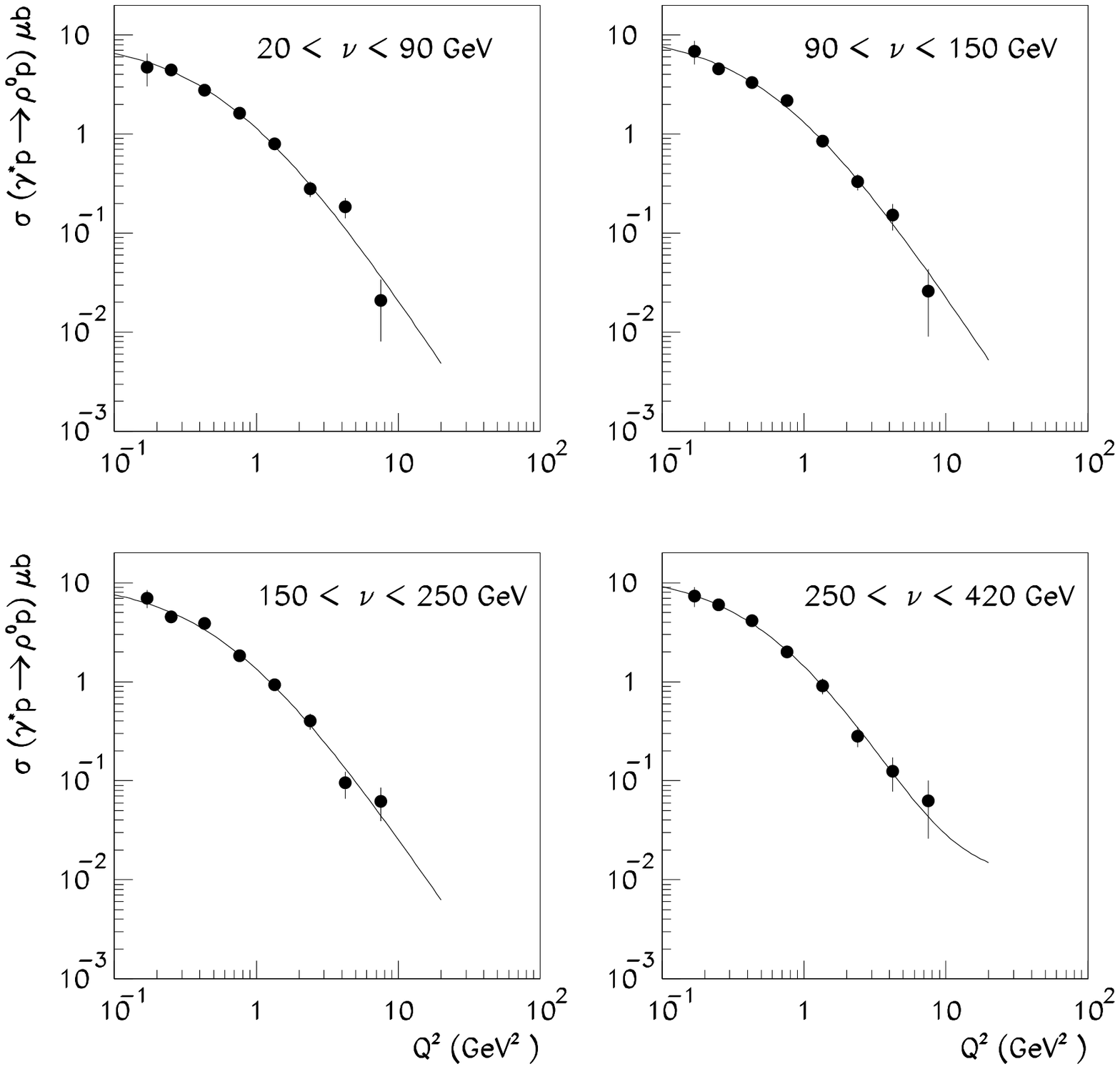}}}

\vskip 2cm

\caption{ The $Q^2$-dependence of cross section $ \sigma _{\gamma ^* p \rightarrow
 \rho ^0 p} (Q^2) $  ($W = 15$ $GeV$ in the left picture, $W = 10.1,$ $14.9,$ $19.3,$
$24.1$ $GeV$ in the right one).}
\vskip 1cm
\end{figure}
As is noticed in the paper~[8], unitarity corrections contribute
to the cross section less than $10\%$, so with present experimental accuracy
we decided to consider the Born term only. It is seen that the Regge method is relevant
enough for the processes (3) description and hence there is no need for taking into account new
poles (probably of non-Regge type) besides well known $\Bbb{P}$ and $\Bbb{R}$. 

Now let us consider the total cross section for the DIS process
$$
\gamma^*+p\to X.
$$
The result of using formula similar to (2) is shown in fig.4. It is obvious that in this case
poles $\P$ and $\R$ are insufficient. Since

\begin{figure}
\vskip  -1cm

\hskip 0.2cm \vbox to 7cm {\hbox to 7cm{\epsfxsize=7cm\epsffile{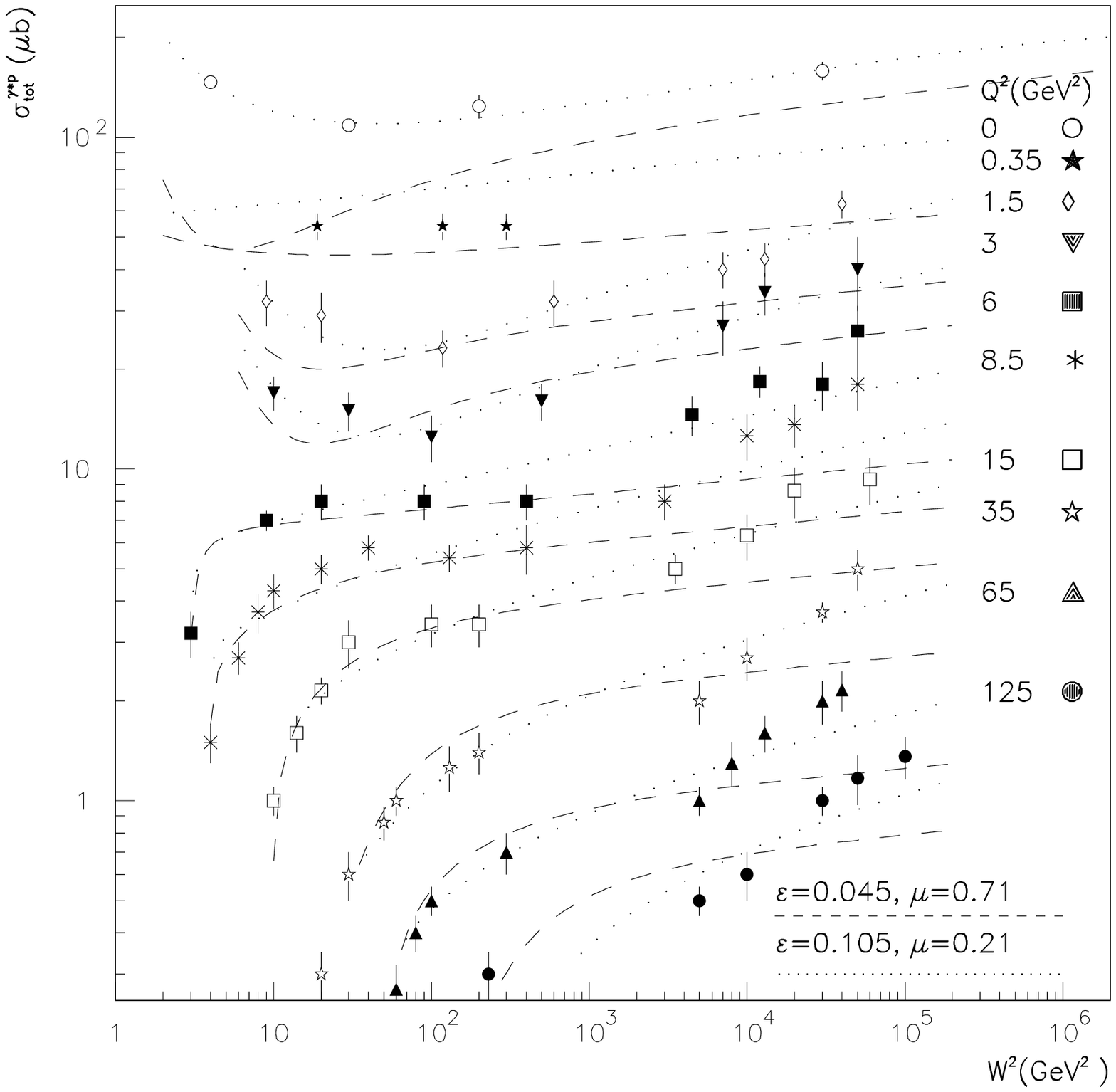}}}

\vskip -7cm
\hskip 8.5cm \vbox to 7cm {\hbox to 7cm{\epsfxsize=7cm\epsffile{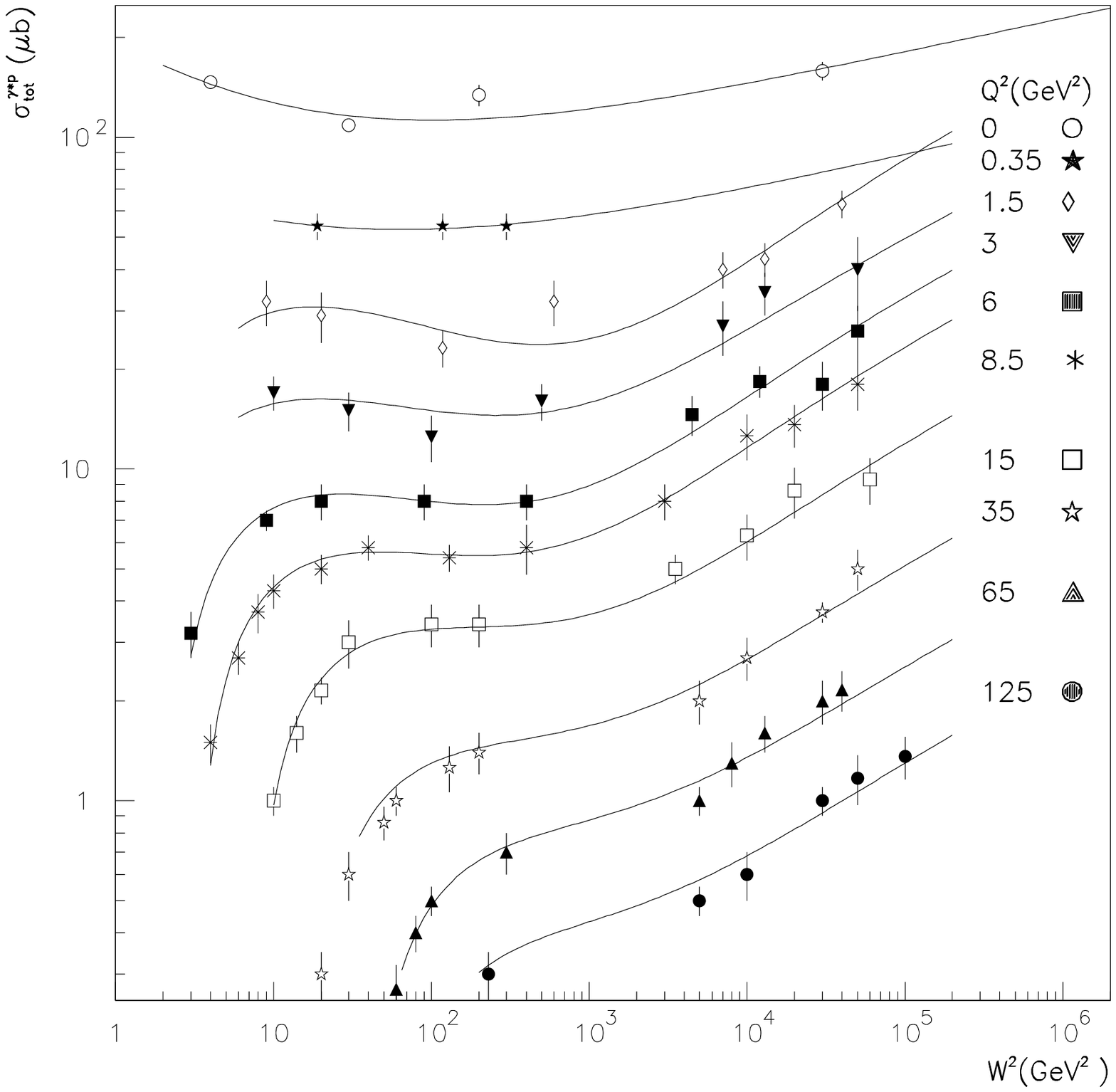}}}
\vskip 2cm
\caption{ The total cross section $ \sigma _{\gamma ^* p \rightarrow
 X} (s=W^2$ $GeV^2,Q^2) $  (the left picture shows the result of using
the two pole model, the right one shows that of using the four pole model).}
\vskip 1cm
\end{figure}

$$
\sigma^{\gamma^*p}_{\mbox\small{{tot}}}=\sum_{V}\sigma_{\gamma^*p\to Vp}+
\sigma^{\gamma^*p}_{\mbox\small{{inel}}},
$$
the insufficiency of ``usual'' Regge poles stems from behaviour of $\sigma^{\gamma^*p}_{\mbox\small{{inel}}}$
which is connected with processes of multiple production.

Does it mean that for successful description of the total cross-sections we
need poles with a high intercept? Not at all. If one adds
``secondary'' Regge poles with $\alpha (0)<1$ then one will have a good description
as is seen in fig.4.They have the following intercepts: 

$$
\alpha_{{\Bbb{R}}_{1}}(0)-1=-0.08;
$$
$$
\alpha_{{\Bbb{R}}_{2}}(0)-1=-0.1.
$$

The problem of interpretation of these new poles, which have intercepts lying close
to each other and to 1, rises immediately. One should also include these poles
into the fit of exclusive cross-sections.

Thus, we have demonstrated that Regge poles describing hadron-hadron processes
do cope with a rapid growth of exclusive vector meson production cross sections
at HERA and we explain this growth as a "threshold effect". Meanwhile it is 
not possible to do the same for the total cross sections unless one adds new poles,
which are not "hard" however.

The authors would like to express their acknowledgment to A.De Roeck kindly 
granted them with experimental data and to J.Field for useful discussions.

\end{document}